\documentclass[citeautoscript,prx,aps,twocolumn,showpacs,showkeys,superscriptaddress,amsmath,amssymb]{revtex4}

\usepackage{graphicx}
\usepackage{psfrag}
\usepackage{epsfig}
\usepackage{color}
\usepackage{subfigure}
\definecolor{dred}{rgb}{0.7,0.0,0.0}
\usepackage{dcolumn}
\usepackage{bm}

\begin{document}

%
%
\title{Phenomenological Three-Orbital Spin-Fermion Model for Cuprates}
%
%
%
\author{Mostafa Sherif Derbala Aly Hussein}
\affiliation{Department of Physics and Astronomy, University of Tennessee,
Knoxville, TN 37966, USA} 
\affiliation{Materials Science and Technology Division,
Oak Ridge National Laboratory, Oak Ridge, TN 37831, USA}


\author{Maria Daghofer}
\affiliation{Institut  f\"ur Funktionelle Materie und Quantentechnologien,
Universit\"at Stuttgart, Pfaffenwaldring 57, D-70569 Stuttgart, Germany} 
\affiliation{Center for Integrated Quantum Science and Technology, University of Stuttgart,
Pfaffenwaldring 57, D-70550 Stuttgart, Germany}

\author{Elbio Dagotto}
\affiliation{Department of Physics and Astronomy, University of Tennessee,
Knoxville, TN 37966, USA} 
\affiliation{Materials Science and Technology Division,
Oak Ridge National Laboratory, Oak Ridge, TN 37831, USA}

\author{Adriana Moreo}
\affiliation{Department of Physics and Astronomy, University of Tennessee,
Knoxville, TN 37966, USA} 
\affiliation{Materials Science and Technology Division,
Oak Ridge National Laboratory, Oak Ridge, TN 37831, USA}

\date{\today}

\begin{abstract}
{A spin-fermion model that captures the charge-transfer properties 
of Cu-based high critical temperature superconductors is introduced and 
studied via Monte Carlo simulations. The strong
Coulomb repulsion among $d$-electrons in the Cu orbitals is phenomenologically 
replaced by an exchange coupling between the spins of the itinerant electrons
and localized spins at the Cu sites, formally
similar to double-exchange models for manganites. 
This interaction induces a charge-transfer insulator gap in the undoped case (five electrons per unit cell). 
Adding a small antiferromagnetic Heisenberg coupling between localized spins 
reinforces the global tendency towards antiferromagnetic order. 
To perform numerical calculations the localized spins are considered classical, as in previous related efforts.
In this first study, undoped and doped $8\times 8$ clusters are analyzed in a 
wide range of temperatures.
The numerical results reproduce experimental 
features in the one-particle spectral function and the density-of-states such as 
{\it (i)} the formation of a Zhang-Rice-like band with a dispersion of 
order $\sim 0.5$~eV  and with rotational symmetry about wavevector $(\pi/2,\pi/2)$ 
at the top of the band, and
{\it (ii)} the opening of a pseudogap at the chemical potential upon doping. We also observed
incipient tendencies towards spin incommensurability.
This simple model offers a formalism intermediate between
standard mean-field approximations, that fail at finite temperatures in regimes with short-range order,
and sophisticated many-body techniques such as Quantum Monte Carlo, that suffer sign problems.}

\end{abstract}
 

 
\maketitle

\section{Introduction} 

The properties of transition metal oxides (TMOs) are determined by two groups of electrons: the $d$-electrons at the transition metals and the $p$-electrons
at the oxygens~\cite{tmo}. The $d$-electrons are believed to be localized and subject 
to strong on-site Coulomb repulsion $U_d$ while the $p$-electrons are considered itinerant with a 
smaller Coulomb repulsion $U_p$. However, the $d$-electrons can be delocalized 
by hybridization with the $p$-electrons and, thus, the degree of hybridization, that varies with
the ratio of the Coulomb repulsion to hopping amplitudes, plays an 
important role in determining the properties of TMOs~\cite{tmo}. 
In addition, the on-site energies $\epsilon_p$ and $\epsilon_d$ of the $p$ and $d$ 
orbitals also affect 
the properties of TMOs~\cite{zaanen}. 
Depending on the relative value of $\Delta=\epsilon_d-\epsilon_p$, $U_d$, and the bandwidth $W$ of the 
itinerant electrons, the latter as determined from the limit when Coulomb repulsion is turned off, 
the TMOs may be in various different regimes. 
The Mott-Hubbard regime occurs when $U_d<\Delta$ and an insulating gap opens in the $d$-band 
if it is half-filled. If $U_d>\Delta$ the system is considered to be in the charge-transfer (CT) 
regime. A gap defined by an electron filled $p$-band and an empty $d$-band opens when 
the $d$-band is nominally half-filled. Systems with large Hubbard repulsions but with $\Delta<W/2$ 
can be metallic~\cite{zaanen}. Recently, even the case of negative 
charge-transfer gaps $\Delta<0$ has been considered~\cite{khomskii,demedici,fratino,bisogni}. 

Among the most important families of TMOs are the high critical 
temperature superconducting cuprates. Their parent compounds are charge-transfer 
insulators (CTI)~\cite{zaanen,elbio}, but from the theory perspective 
they have been studied primarily using single-orbital Hubbard or $t-J$ models because these models are simpler
than more realistic multiorbital Hamiltonians that include oxygens. 
Using simplified one-band models is justified
by the experimental observation of a single-band Fermi surface~\cite{shen,allen,wells,damas} 
and also by the Zhang-Rice singlet concept where the three-orbital 
Hubbard model is approximately mapped into an effective $t-J$ model~\cite{ZR}. 
While many properties of the cuprates have been 
captured by single-band models~\cite{elbio}, several questions regarding the 
role of the oxygen remain. 
One of the main issues are the differences between the properties of doped Mott insulators, 
described by single-band models, and 
charge-transfer insulators where both the $d_{x^2-y^2}$ Cu orbital and 
the $p_{\sigma}$ O orbitals are considered.
Early numerical studies of three-band models did not indicate 
major physical differences among both approaches~\cite{hybertsen,bacci,maria}, 
but other authors have claimed that the 
multiorbital character plays a crucial role in the physics of the cuprates~\cite{emery,sawatzky}. 

The discovery of the iron-based superconductors~\cite{kamihara,chen1,chen2,wen} 
brought to the forefront the need to develop models and numerical 
approaches to deal with multi-orbital systems. In this context, effective multi-orbital 
spin-fermion models were developed that allowed the study of 
many properties of these materials such as magnetic phases, density of states, 
Fermi surface, and resistivity, among others~\cite{shuhua13}. These efforts on iron pnictides
and chalcogenides actually built upon the double-exchange models for manganites. The 
aim of the present work is to develop a spin-fermion model for the CuO$_2$ planes 
of the cuprates that can be studied with the Monte Carlo techniques previously 
developed for the pnictides, with the goal to understand, at least qualitatively, 
the role played by the O $p_{\sigma}$-orbitals.  

A single-orbital, as opposed to multi-orbital, spin-fermion model for the cuprates 
was developed in the 90s~\cite{opstripes}. In that early effort, the Cu $d$-band 
was split via a spin-spin interaction among phenomenological localized spins 
and the spins of the intinerant electrons, similarly as in the model proposed here. 
This interaction prevents the double 
occupancy of the Cu sites, crudely mimicking 
the Hubbard on-site repulsion effects. By using classical localized spins and Monte Carlo,
several of the static and dynamical properties of the cuprates were reproduced showing that
this avenue, that interpolates between traditional mean-field approximations and far more complicated Quantum
Monte Carlo approaches, is fruitful. Magnetic incommensurability and a short-distance 
tendency towards $d$-wave pairing was observed upon doping~\cite{opstripes,opdwave}. 

In the present effort, starting with the standard tight-binding term of the three-orbital 
Hubbard model for cuprates and introducing phenomenological localized spins, we will find 
the interaction parameters values that better reproduce the density of states (DOS) 
of the full Cu oxide Hamiltonian.
The tight-binding term involves 3$d_{x^2-y^2}$ Cu and 2$p_{\sigma}$ (2$p_x$ or 2$p_y$) 
orbitals of the two oxygens in the CuO$_2$ unit cell. As already explained,
the Cu-sites Hubbard repulsion that splits the half-filled $d$-band will be 
replaced by a magnetic coupling between 
the spin of the itinerant electrons when at the $d$-orbital and 
Cu localized spins.
A small antiferromagnetic Heisenberg coupling among nearest-neighbor localized spins 
enhances the global antiferromagnetic tendencies. 
In addition, the spins of the $p$-orbital electrons  
are coupled antiferromagnetically to their two neighboring localized spins. 

In the undoped case, with five electrons per CuO$_2$ unit cell, it will be shown 
that the model leads to a charge-transfer insulator where, unexpectedly, 
the gap states have approximately {\it equal} amounts of $p$ and $d$ character. This is contrary
to the widely held perception that holes reside primarily at the oxygens.
This is also different from the one-orbital Mott insulator approach 
in which one-single orbital contributes entirely to the states that define the gap. 

Several other interesting results were obtained. For instance, 
long-range antiferromagnetic order, as in the parent compound of the cuprates, 
develops with reducing temperature. Incipient tendencies towards spin incommensurability
were observed with doping.
Even more importantly, a study of the one-particle spectral functions indicates that, 
in agreement with angular-resolved photoemission (ARPES) results for the undoped cuprates, 
states with wavevectors $(\pm\pi/2,\pm\pi/2)$ are the first to accept doped holes. The ARPES region
around these wavevectors are rotationally symmetric, with equal curvature in all directions,
a feature reproduced in single-band models only after the addition of longer range 
hoppings, while in our approach it emerges spontaneously without fine tuning.
In addition, the lowest state for electron doping has momentum $(\pi,0)$ and $(0,\pi)$, as expected.
Moreover, 
a Zhang-Rice-like singlet (ZRS) band spontaneously appears in the DOS, and
a pseudogap at the chemical potential develops upon doping.

The paper is organized as follows: in Section~\ref{3bhm} results for the DOS of 
the full undoped three-orbital Hubbard model are presented to guide the tuning of parameters 
in the proposed spin-fermion model which is introduced in Section~\ref{model}. 
Results for the DOS and the one-particle spectral functions (photoemission), as well as the 
magnetic structure factor, 
are presented in Section~\ref{results}, while Section~\ref{conclu} is devoted to the conclusions.

\section {Charge-Transfer Regimes in the Three-Band Hubbard Model}\label{3bhm}

\begin{figure}[thbp]
\begin{center}
\includegraphics[width=7cm,clip,angle=0]{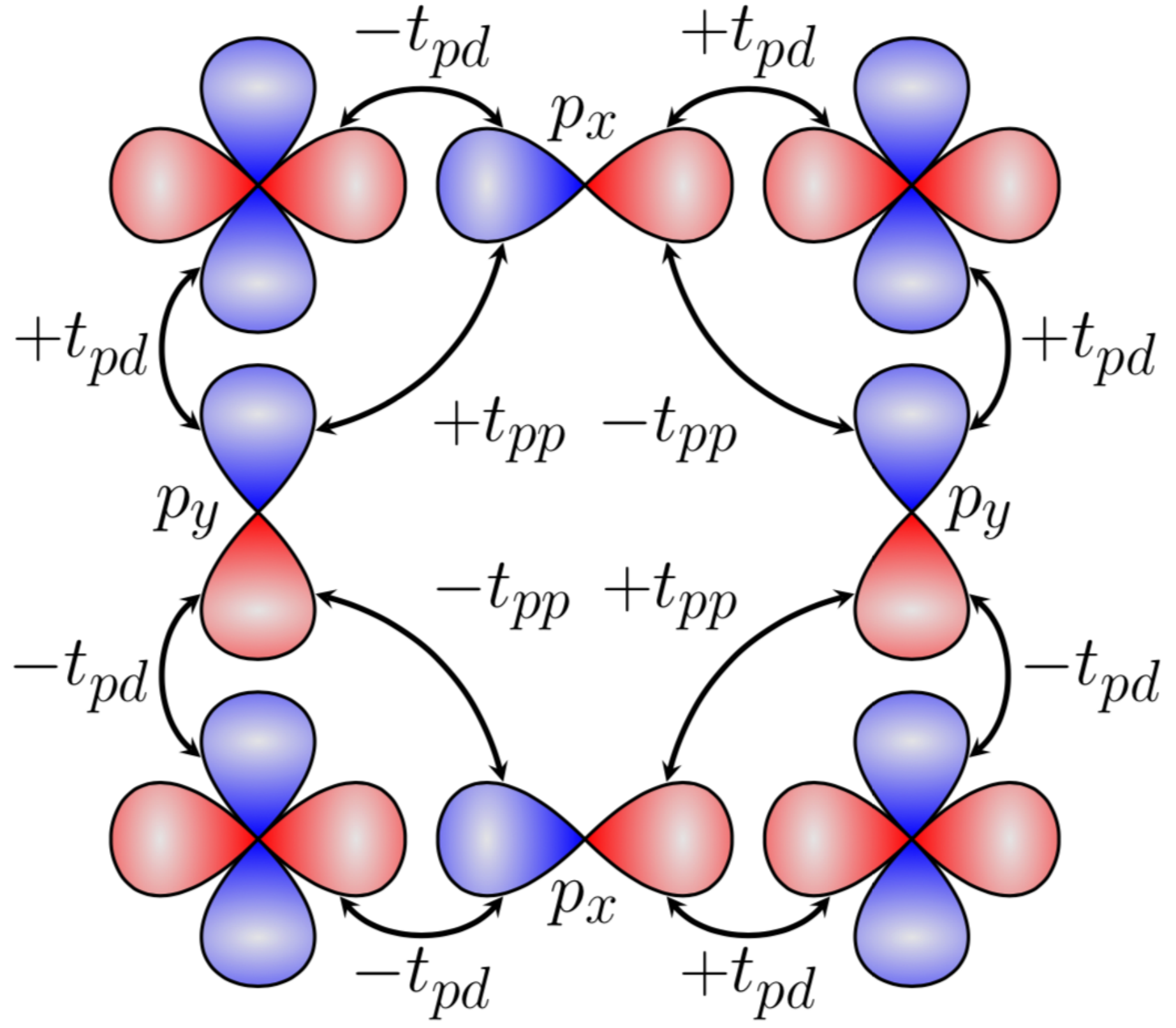}
\vskip -0.3cm
\caption{(color online) Schematic drawing of the Cu $d_{x^2-y^2}$ orbitals at the copper sites of the square lattice,
with the sign convention indicated by the colors (red for + and blue for -). The oxygen $p_{\sigma}$ orbitals 
with their corresponding sign convention are also shown, located at the Cu-O-Cu bonds.
The sign convention for the $t_{pd}$ and $t_{pp}$ hoppings is also indicated.} 
\vskip -0.4cm
\label{cuo2fig}
\end{center}
\end{figure}

The band gaps and electronic structures of TMOs were described before~\cite{zaanen} 
and they depend on the relationship between the charge-transfer energy $\Delta$ and
the $d$-$d$ Hubbard repulsion $U_d$. In general, if $U_d<\Delta$ the band gap 
of the undoped state is controlled by $U_d$ and the system 
is a Mott-Hubbard insulator, while if $U_d>\Delta$ the gap is 
controlled by $\Delta$ and of charge-transfer nature for $\Delta>W/2$, 
where $W$ is the bandwidth of the oxygen $p$-band~\cite{zaanen}. 
However, the role of the hybridization
between the $d$ and $p$ bands, important in cuprates, 
is often neglected. For this reason, first we present results for the 
orbital-resolved density-of-states of the undoped 
three-orbital Hubbard model obtained using the variational cluster approach (VCA)~\cite{maria}. 
The Hamiltonian, in electron notation, is given by

\begin{equation}
H_{\rm 3BH}= H_{\rm TB}+ H_{\rm int},
\label{3bh}
\end{equation}
\noindent where 
\begin{equation}\begin{split}
H_{\rm TB} = -t_{pd}\sum_{{\bf i},\mu,\sigma}\alpha_{{\bf i},\mu}(p^{\dagger}_{{\bf i}+{\hat\mu\over{2}},\mu,\sigma}d_{{\bf i},\sigma}+ h.c.)-\\
t_{pp}\sum_{{\bf i},\langle\mu,\nu\rangle,\sigma}\alpha'_{{\bf i},\mu,\nu}[p^{\dagger}_{{\bf i}+{\hat\mu\over{2}},\mu,\sigma}(p_{{\bf i}+{\hat\nu\over{2}},\nu,\sigma}+p_{{\bf i}-{\hat\nu\over{2}},\nu,\sigma})+ h.c.]\\
+\epsilon_d\sum_{{\bf i}}n^d_{{\bf i}}+\epsilon_p\sum_{{\bf i},\mu}n^p_{{\bf i}+{\hat\mu\over{2}}}+\mu_e\sum_{{\bf i},\mu}(n^p_{{\bf i}+{\hat\mu\over{2}}}+n^d_{\bf i}),
\label{htb}
\end{split}\end{equation}
\noindent and
\begin{equation}
\begin{split}
H_{\rm int} = U_{d}\sum_{{\bf i}}n^d_{\bf i,\uparrow}n^d_{\bf i,\downarrow}
+U_{p}\sum_{{\bf i},\mu,\sigma}n^p_{{\bf i}+{\hat\mu\over{2}},\uparrow}n^p_{{\bf i}+{\hat\mu\over{2}},\downarrow}.
\label{int}
\end{split}
\end{equation}
\noindent The operator $d^{\dagger}_{{\bf i},\sigma}$ creates an electron with spin $\sigma$ 
at site ${\bf i}$ of the copper square lattice, while
$p^{\dagger}_{{\bf i}+{\hat\mu\over{2}},\mu,\sigma}$ creates an electron with spin $\sigma$ 
at orbital $p_{\mu}$, where $\mu=x$ or $y$, for the oxygen located at
${\bf i}+{\hat\mu\over{2}}$. The hopping amplitudes $t_{pd}$ and $t_{pp}$ 
correspond to the hybridizations between nearest-neighbors 
Cu-O and O-O, respectively, and $\langle\mu,\nu\rangle$ indicate O-O pairs connected by $t_{pp}$ as indicated in
Fig.~\ref{cuo2fig}. $n^p_{{\bf i}+{\hat\mu\over{2}},\sigma}$ ($n^d_{{\bf i},\sigma}$) 
is the number operator for $p$ ($d$) electrons with spin $\sigma$, and
$\epsilon_d$ and $\epsilon_p$ are the on-site energies at the Cu and O sites, respectively. 
If $\epsilon_d=0$ then $\Delta=\epsilon_d$ -$\epsilon_p$,
is the charge-transfer gap. The Coulomb repulsion between two electrons at the same site and 
orbital is $U_d$ ($U_p$) for $d$ ($p$) orbitals. The signs
of the Cu-O and O-O hoppings due to the symmetries of the orbitals is included in 
the parameters $\alpha_{{\bf i},\mu}$ and $\alpha'_{{\bf i},\mu,\nu}$ and follows 
the convention shown in Fig.~\ref{cuo2fig}. Finally, $\mu_e$ is the electron chemical potential.

The orbital-resolved DOS in the electron representation for the accepted values of $U_d=8t$~\cite{maria,hybertsen} 
and $U_p=3t$ (where $t=t_{pd}$ is the energy unit) is in Fig.~\ref{3bh}~(a). The effect of the Coulomb repulsion on the DOS can be 
understood by comparing with the tight-binding band dispersion in Fig.~\ref{TBdisp}. The 
spectral weight associated to the portion of the band above the chemical potential in  Fig.~\ref{TBdisp} 
appears to the right of the chemical potential in Fig.~\ref{3bh}. It is clear that the gap where 
the chemical potential is located in Fig.~\ref{3bh} results mostly from the split, due to $U_d$, 
of the top band in Fig.~\ref{TBdisp} which, as shown in the 
figure, arises mostly from the $d$ orbitals (that in the electron picture are on top). 
In the non-interacting limit, this band has a small oxygen content due to the hybridization 
$t_{pd}$ and it has a similar dispersion to the tight-binding band of the single-band 
Hubbard model when $t'=-0.3t$ and $t''=0.2t$ hoppings are added~\cite{senechal,maria,macridin}. 
Thus, the gap opening in the top band is ``captured'' by the single-particle model with $U=8t$~\cite{maria}. 

As shown in Fig.~\ref{3bh}~(a), the charge-transfer gap where the chemical potential resides at
$U_d/t=8$ is about $2t$, {\it similar} in magnitude to the Mott gap of the single-band Hubbard model 
with $U/t=8$\cite{elbio,ortolani}. Naively, the DOS gap would be expected to be proportional to $U$, 
but in both cases screening effects reduce the gap. The main qualitative difference, though, 
lies in the orbital composition of the band. As shown in Fig.~\ref{3bh}~(a), the spectral weight occupied 
by electrons immediately at the left of the chemical potential has a 50\%-50\% $p-d$ orbital composition
indicating its charge-transfer character (red and blue curves have almost identical weight). 
This is due to the additional hybridization effects 
caused by the strong Coulomb interaction that affects the spectral weight from the 
$p$ tight-binding bands. On the other hand, the spectral weight at the 
right of the chemical potential in Fig.~\ref{3bh}~(a) 
is mostly of $d$-character i.e. when electrons 
are added they populate $d$-Cu orbitals.

\begin{figure}[thbp]
\begin{center}
\includegraphics[width=8cm,clip,angle=0]{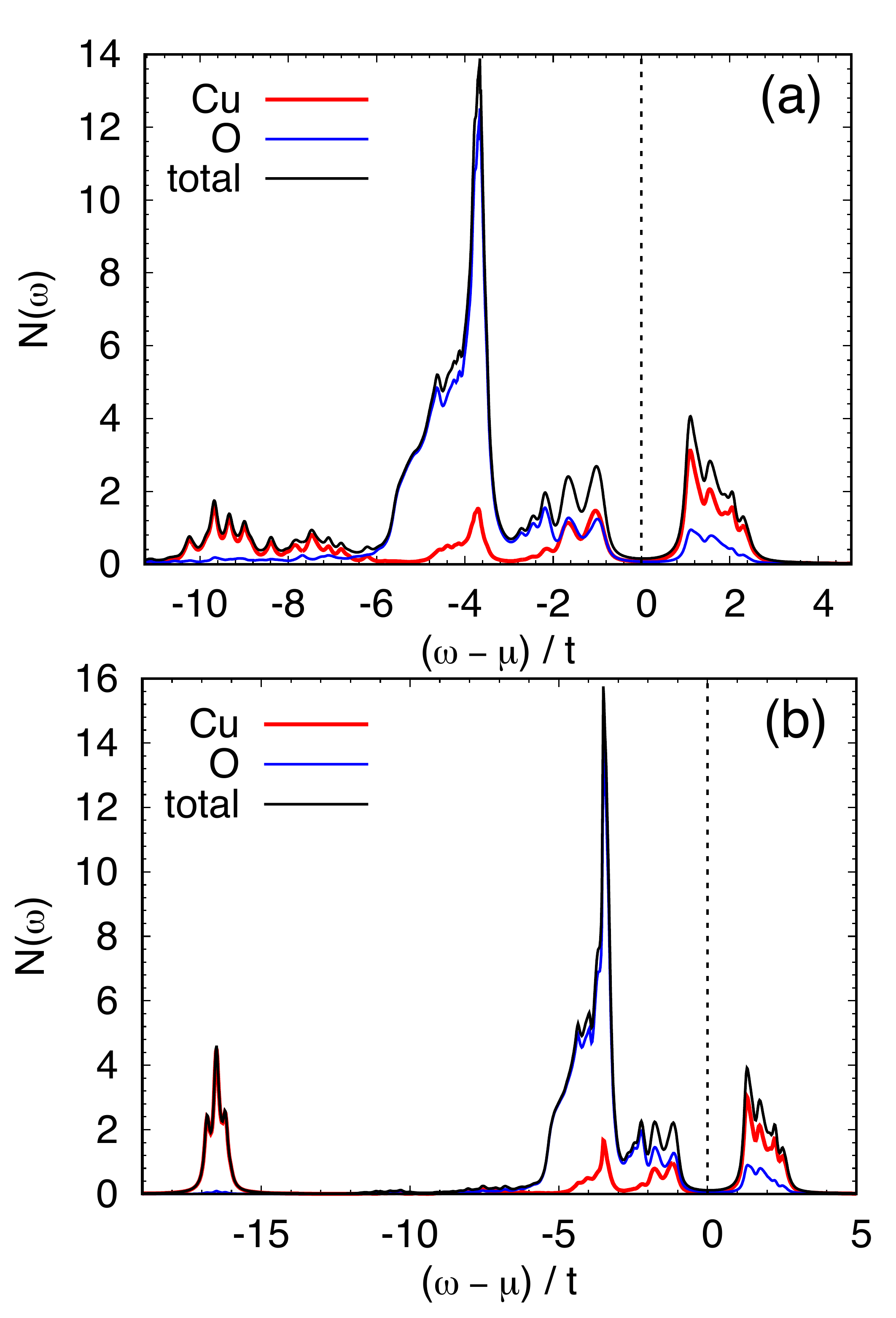}
\vskip -0.3cm
\caption{(color online) Orbital-resolved density-of-states for the full three-band Hubbard model 
with the parameters used before in Ref.~\cite{maria} where $t_{pd}=t$ is the unit of energy and
$t_{pp}=0.5t$, $\Delta=3t$, and $U_p=3t$. Panel (a) corresponds to $U_d=8t$ while 
panel (b) to $U_d=16t$. The dashed line indicates the chemical potential in the undoped
case with one hole (five electrons) per CuO$_2$ unit cell. Results are shown in the electron notation.} 
\vskip -0.4cm
\label{3bh}
\end{center}
\end{figure}

Note also that a small amount of spectral weight, 
almost 100\% of $d$ character, has been transfered to lower 
energy (in the electron picture) in the 
interval $-10<(\omega-\mu)/t<-6$. This weight was previously identified by 
some authors as the ``lower Hubbard band'' (LHB)~\cite{demedici} although a well-defined LHB 
is not sharply developed at the value of $U_d$ considered realistic. 
In fact, we found that to develop a well-defined LHB, as in an extreme 
charge-transfer system~\cite{zaanen}, a $U_d$ as large as $16t$ is required. 
The DOS in this situation is in Fig.~\ref{3bh}~(b). The LHB is located at $(\omega-\mu)/t\approx -16$ and
it has 100\% $d$ character. Now the separation between the upper and lower Hubbard bands is approximately 
$U_d$ while the charge-transfer gap is only slightly reduced. 
Still for the two values of $U_d$ presented in Fig.~\ref{3bh} it is clear that due to the $p-d$ 
hybridization, arising from the combined effect of interorbital hopping and Coulomb interaction, 
the states that define the charge-transfer gap have {\it mixed} orbital character~\cite{nohyb}. 
This indicates that doped holes will go both into the 
oxygens {\it and} the coppers since the spectral weight is comparable among $p$ and $d$ orbitals. 
In summary, the deviations clarified in this section 
from the simplistic view of either purely Hubbard or purely charge-transfer gap materials were
not emphasized before in the literature, increases the level of complexity of the system, and 
will be an important feature that we will try to capture in the effective model presented next. 
We conclude this section stating that cuprates
are not sharply ``charge-transfer'' insulators but they reside at the intersection between the Hubbard and
charge-transfer families.

\section {Effective Three-Band Model for CuO$_2$ Planes}\label{model}

The starting point for the effective model that we will develop is the tight-binding portion 
of the three-band Hubbard model given in Eq.~\ref{htb} with $t_{pd}=1.3$~eV and $t_{pp}=0.65$~eV, 
on-site energy $\epsilon_p=-3.6$~eV~\cite{hybertsen}, and a
$\Delta=\epsilon_d-\epsilon_p$ which is positive ($\epsilon_d=0$)~\cite{negdel}. 

\begin{figure}[thbp]
\begin{center}
\includegraphics[width=8.5cm,clip,angle=0]{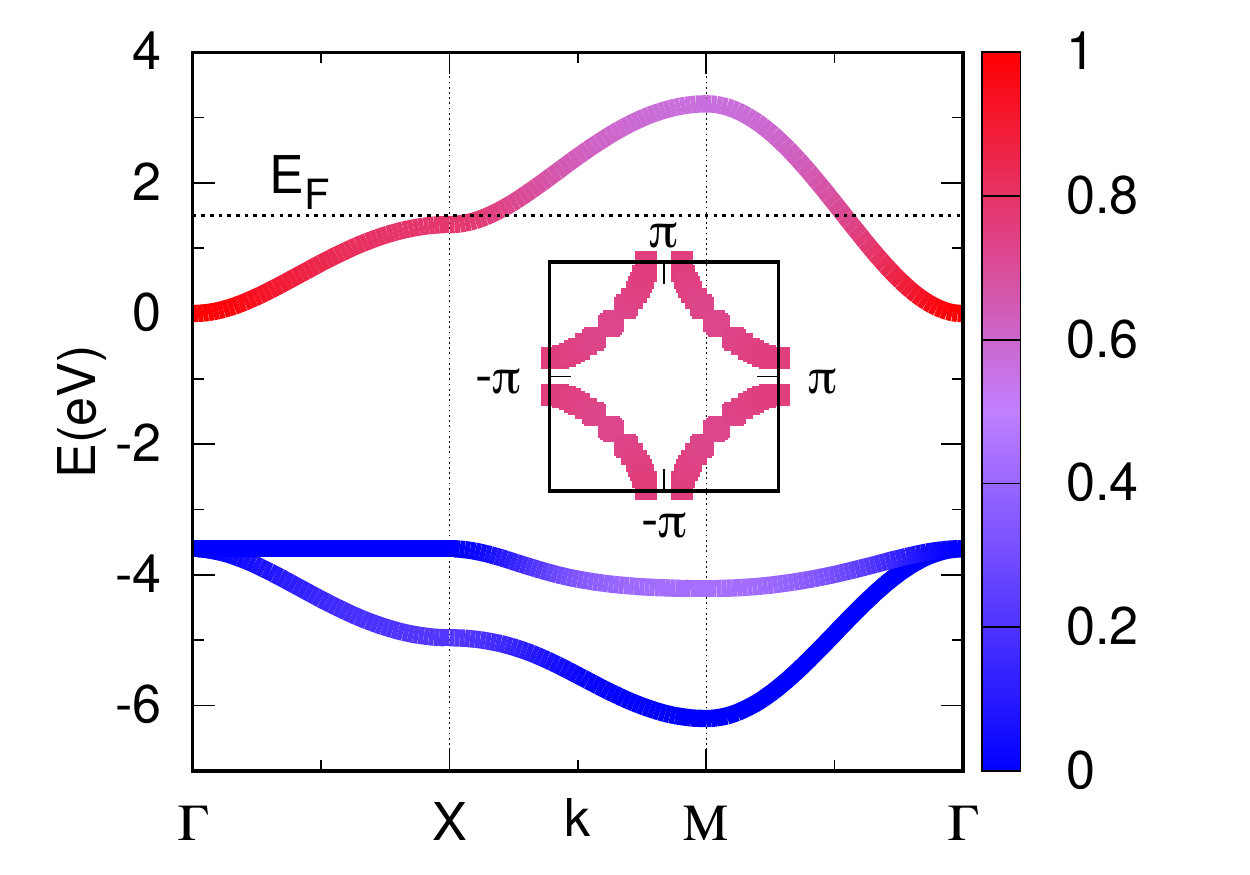}
\vskip -0.3cm
\caption{(color online) Band dispersion for the tight-binding term of the CuO$_2$ Hamiltonian. 
The orbital content is displayed with red (blue) indicating $d$ ($p$) character. The dashed
line indicates the position of the chemical potential (or Fermi level E$_{\rm F}$) 
at density $\langle n \rangle=5$ (undoped case). 
The Fermi surface at this density is in the inset. Colors indicate the 
orbital content of the bands, with
the palette on the right denoting the weight of the $d$ component (e.g. 1 means 
100\% copper $d$, and the oxygen weight is simply one minus the copper weight).} 
\vskip -0.4cm
\label{TBdisp}
\end{center}
\end{figure}

Note that in the electron representation the undoped case is characterized by 
one hole at the coppers and no holes at the oxygens, which 
corresponds to five electrons per CuO$_2$ unit cell (the maximum possible electronic 
number in three orbitals is six). The orbital-resolved tight-binding bands along the 
$\Gamma-X-M-\Gamma$ path in the Brillouin zone calculated on a $100\times 100$ square lattice 
(with coppers at the sites of the lattice) is in Fig.~\ref{TBdisp}. The dashed black line is 
the chemical potential for electronic density $\langle n\rangle=5$ and the 
corresponding Fermi surface is in the inset. An analysis of the orbital composition 
of each of the three bands, shown by the color palette in the figure, indicates that the 
top band is purely $d$ at the $\Gamma$ point and becomes hybridized with the $p$ orbitals so that its $d$ content 
is 78\% at X and 56\% at M. The two bottom bands have pure $p$ 
character at the Brillouin zone center. The middle band achieves 43\% $d$ character at M, while 
the lower band has 21\% $d$ character at X. Note that the
tight-binding Fermi surface, shown in the inset, has the qualitative form expected 
in the cuprates. However, its orbital content is about 75\% $d$ only, showing that the oxygen component
is not negligible even if only one band crosses the Fermi level.
 
The interaction term in the spin-fermion model is purely {\it phenomenological}, as in all spin-fermion models in previous
literature. It is introduced to prevent double occupancy in the $d$ orbitals by creating lower and 
upper bands, while spectral weight originating in the $p$ orbitals remains in the middle, 
in such a way that a charge-transfer insulator results for 
five electrons per unit cell. To achieve these goals, we introduce phenomenological localized spins at the 
Cu sites. These on-site spins will be coupled via an antiferromagnetic 
coupling $J_{\rm Sd}$ to the spins of the mobile $d$-electrons at the same site via
\begin{equation}
H_{\rm Sd} = J_{\rm Sd}\sum_{{\bf i}}{\bf S_i.s_i},
\label{hSd}
\end{equation}
\noindent where ${\bf S_i}$ denotes the localized spins at ${\bf i}$, 
${\bf s_i}=d^{\dagger}_{{\bf i},\alpha}\vec\sigma_{\alpha\beta}d_{{\bf i},\beta}$ 
is the spin of the mobile $d$-electrons, and $\vec\sigma_{\alpha\beta}$ are Pauli matrices. 
Since this term is phenomenological, in principle the coupling between localized and 
itinerant spins can be either anti- (AF) or ferromagnetic (FM) since in the AF (FM) case the 
lower $d$-band will contain electrons with spins 
antiparallel (parallel) to the localized spins. For the classical localized spins used here,  
the results are independent of the sign of $J_{\rm Sd}$ and we will simply 
consider the AF coupling as our convention. Note that in the 
absence of electronic hopping this interaction would lead to a half-filled $d$-band and 
totally filled $p$-bands for the overall density $\langle n\rangle=5$ per CuO$_2$ cell.

To enhance further the tendency towards antiferromagnetic order in the undoped case, 
as in real undoped cuprates, an 
antiferromagnetic Heisenberg coupling $J_{\rm AF}$ between the localized spins is also introduced via 
\begin{equation}
H_{\rm AF} = J_{\rm AF}\sum_{{\bf i}}{\bf S_i.S_i}.
\label{hAF}
\end{equation}
Finally, a coupling $J_{\rm Sp}$ between the localized spins and the $p$-electrons spins
at each of the four neighboring oxygens is added (introducing effectively magnetic frustration upon doping)
\begin{equation}
H_{\rm Sp} = J_{\rm Sp}\sum_{{\bf i,\hat\mu}}{\bf S_i.s}_{{\bf i}+{\hat\mu\over{2}}},
\label{hSp}
\end{equation}
\noindent where $\hat\mu=\pm\hat x$ or $\pm\hat y$ and ${\bf s}_{{\bf i}+{\hat\mu\over{2}}}=p^{\dagger}_{{\bf i}+{\hat\mu\over{2}},\mu,\alpha}\vec\sigma_{\alpha\beta}p_{{\bf i}+{\hat\mu\over{2}},\mu,\beta}$.

Thus, the spin-fermion (SF) Hamiltonian defined here is given by four terms as
\begin{equation}
H_{\rm SF} = H_{\rm TB} + H_{\rm Sd} + H_{\rm AF} + H_{\rm Sp}.
\label{ham}
\end{equation}
This phenomenological Hamiltonian we propose is reminiscent of the model in Ref.~\cite{sawatzky} 
except that they work in the limit where the $d$ electrons are fully localized and only
contribute their magnetic degree of freedom.

\begin{figure}[thbp]
\begin{center}
\includegraphics[width=9cm,clip,angle=0]{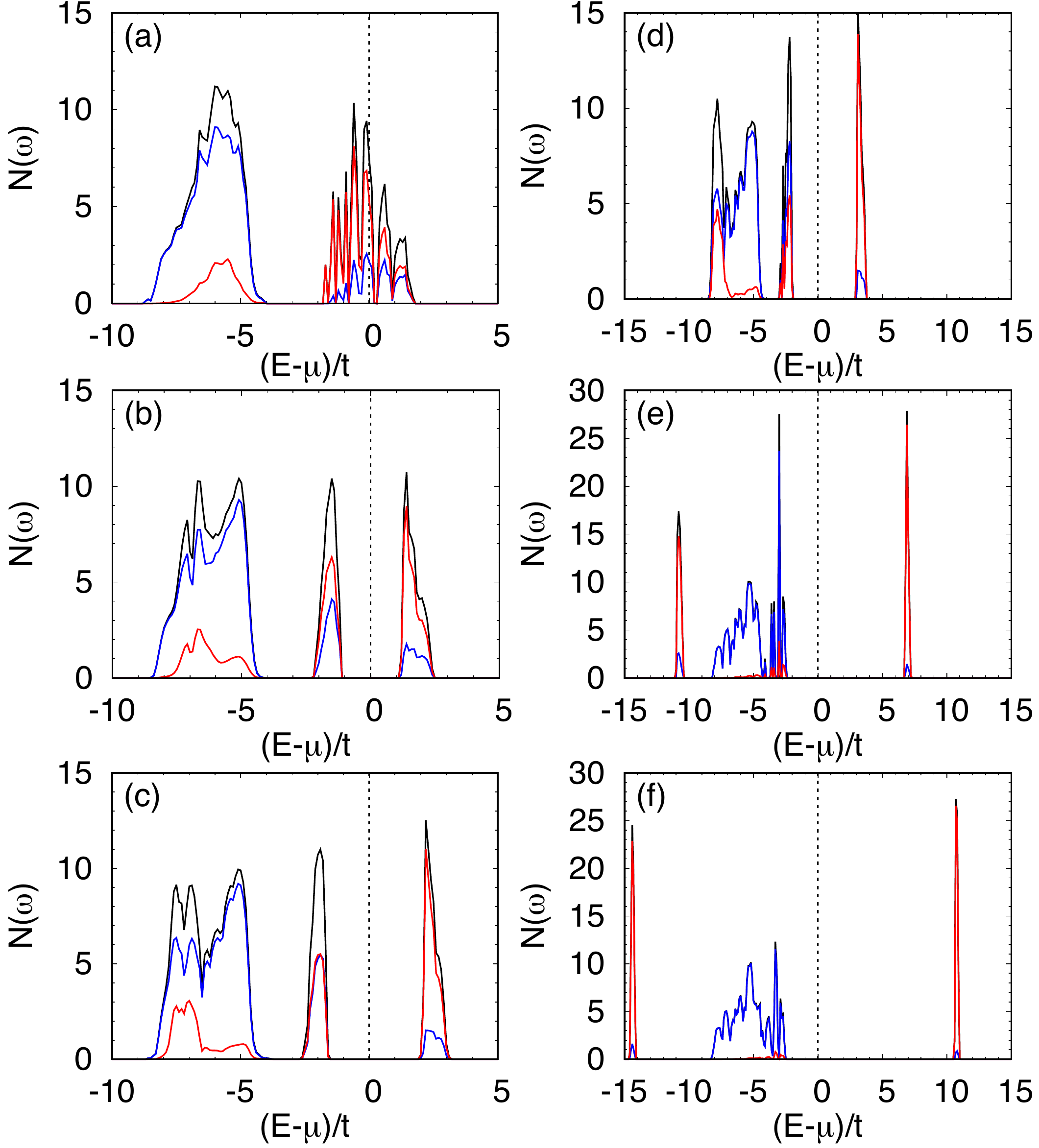}
\vskip -0.3cm
\caption{(color online) 
Orbital-resolved density-of-states for the spin-fermion model 
with $J_{\rm AF}=0.1$~eV and $J_{\rm Sp}=1$~eV. The various panels correspond to (a) $J_{\rm Sd}=0$,
(b) $J_{\rm Sd}=2$, (c) $J_{\rm Sd}=3$, (d) $J_{\rm Sd}=4$, (e) $J_{\rm Sd}=8$, 
and (f) $J_{\rm Sd}=12$ (all in eV units). Results are for the undoped case, i.e. $\langle n\rangle=5$
and were obtained using an $8 \times 8$ lattice at temperature $T \sim 120$~K. 
The $d$ ($p$) spectral weight is in red (blue) while the total spectral
weight is indicated by the black line. The chemical potential is at the vertical dashed line.} 
\vskip -0.4cm
\label{dospos}
\end{center}
\end{figure}

The computational simplification that allows 
the numerical study of our Hamiltonian is that the localized 
spins are assumed classical~\cite{foot}. With this approximation, the full $H_{\rm SF}$ can be studied with 
the same Monte Carlo (MC) procedure widely employed before for 
the pnictides~\cite{shuhua13} and double-exchange 
manganites~\cite{manganites}. 
 
To select the values of the couplings, we studied the properties 
of the model for a variety of parameters finding
the combination that better reproduced some experimental properties of the cuprates and the
results in Fig.~2. 
In Fig.~\ref{dospos}, we present the orbital-resolved density-of-states 
for $J_{\rm AF}=0.1$~eV, $J_{\rm Sp}=1$~eV, and several values of $J_{\rm Sd}$. 
At $J_{\rm Sd}=0$ in panel (a), the chemical potential 
(vertical dashed line) is in the middle of the upper band of mostly $d$-character 
and the system is metallic. However, at $J_{\rm Sd}=2$~eV, panel (b), the upper 
band is split. Now the undoped system is an insulator with the chemical potential inside a gap. 
While the gap is similar to the charge-transfer gap of the 
cuprates $\Delta\approx 2$~eV~\cite{ctgap}, note that the band to the left of $\mu$ 
has primarily $d$-character. 
By increasing further $J_{\rm Sd}$ both the magnitude of the insulating 
gap and the $p$ composition of the band below $\mu$ increases. 
We found that for $J_{\rm Sd}=3$~eV, panel (c), the $d$ and $p$ orbitals contribute {\it equally} 
to the density-of-states just below the chemical potential as in the three-orbital Hubbard model discussed before 
with $U_d=8t$ [Fig.~\ref{3bh}~(a)], and the charge-transfer 
gap is about 3~eV. If $J_{\rm Sd}$ continues to increase, then the $d$ spectral weight 
continues to be redistributed and for $J_{\rm Sd}=4$~eV [panel (d)] there is more $p$ than $d$ weight to the left 
of the chemical potential, but no sharp lower-band has yet developed 
(equivalent to a Hubbard lower-band). This lower band develops when $J_{\rm Sd}=8$~eV as shown in panel (e).
Finally, for extreme values, such as $J_{\rm Sd}=12$~eV in panel (f), the $p-d$ hybridization is removed 
and the upper and lower $d$-bands surround the pure $p$ bands. After this analysis, we set $J_{\rm Sd}=3$~eV as 
the value that may better capture the properties of the cuprates.   

We observed that if the signs of the couplings $J_{\rm Sd}$ and $J_{\rm Sp}$  
are simultaneously reversed, turning both couplings FM, the results are the same except 
that the up and down spins are interchanged since the only modification 
in the Hamiltonian is that $\sigma\rightarrow -\sigma$. However, if only the
sign of one of the couplings is changed, for example $J_{\rm Sd}=-3$~eV, the results 
are different and the system develops phase separation (details not shown). For this reason only
AF couplings between the itinerant and the localized spins will be considered here. 

The calculations shown below 
were performed using squared $8\times 8$ clusters with periodic boundary conditions (PBC). 
These lattice sizes are larger than those accessible to study the
three-band Hubbard model either via quantum Monte Carlo~\cite{muramatsu,gubernatis,johnston} or DMRG~\cite{white}.
During the simulation the localized spins ${\bf S_i}$ evolve via a standard Monte Carlo procedure, 
while the resulting single-particle 
Hamiltonian for the itinerant $p$ and $d$ electrons is exactly diagonalized~\cite{manganites}. 
The present simulations are performed at inverse temperature $\beta=(k_BT)^{-1}$ ranging from
$10$ to $400$ in units of eV$^{-1}$, or temperature T from 1200~K to 30~K~\cite{foot1}. Reaching such low temperatures is an advantage 
of the present approach because for Hubbard model 
quantum Monte Carlo studies can only be performed at high temperatures due to the
``sign problem'' while  DMRG can only be performed at zero temperature and ladder-like cylindrical geometries.

\section {Results}\label{results}


\subsection {Density of States and Band Structure}

\begin{figure}[thbp]
\begin{center}
\includegraphics[width=8cm,clip,angle=0]{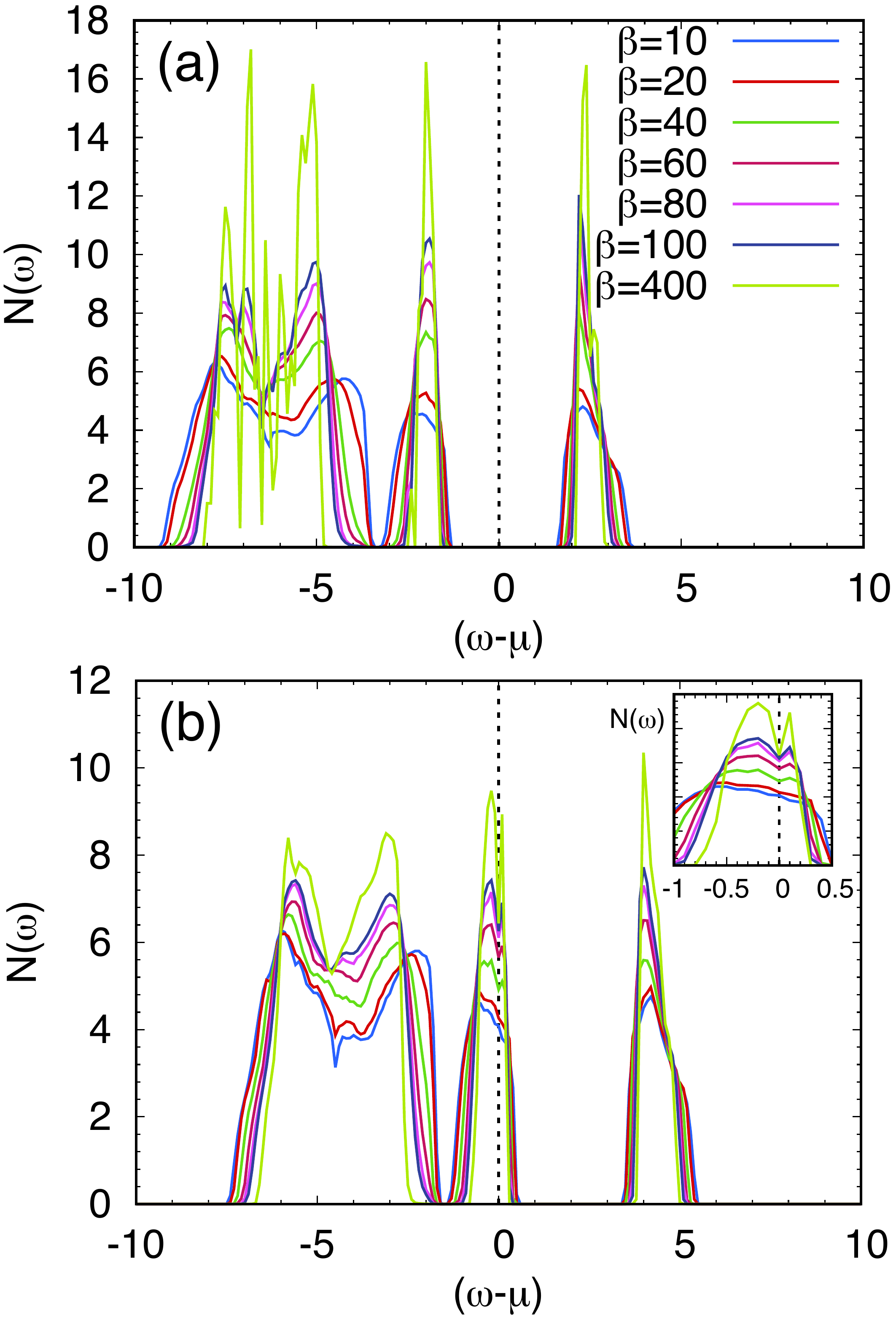}
\vskip -0.3cm
\caption{(color online)  Spin-fermion model density-of-states
with $J_{\rm Sd}$=3, $J_{\rm Sp}$=1, and $J_{\rm AF}$=0.1 (all in eV) using an $8\times 8$ 
lattice and several inverse temperatures ($\beta = 400$ corresponds to $T \sim 30$~K
while $\beta = 10$ to $T \sim 1200$~K).
(a) corresponds to the undoped case 
$\langle n \rangle$=5 while (b) is at 25\% doping $\langle n \rangle=4.75$ (16 holes). 
The inset shows the pseudogap in the ZRS band at the chemical potential.}
\vskip -0.4cm
\label{dosh0h16}
\end{center}
\end{figure}

The DOS for the undoped case ($\langle n\rangle=5$) was calculated for 
$\beta/t$ ranging from 10 to 400 and $t=1$~eV. Because of the $J_{\rm Sd}$ 
interaction, the width of the spectrum
increases from 9.5~eV in the non-interacting case (Fig.~\ref{TBdisp}) to about 12~eV 
at $J_{\rm Sd}=3$~eV [Fig.~\ref{dospos}~(c)] and in this case, as shown in Fig.~\ref{dosh0h16}~(a), 
the chemical potential is in a charge-transfer gap. The dispersion of the bands is reduced as the temperature decreases 
rendering the features in the DOS sharper. In addition, to the left of the chemical potential 
there are two structures, and the peak closest to the chemical potential could be identified 
with a band resembling the Zhang-Rice singlet (ZRS) band.

\begin{figure}[thbp]
\begin{center}
\includegraphics[width=8cm,clip,angle=0]{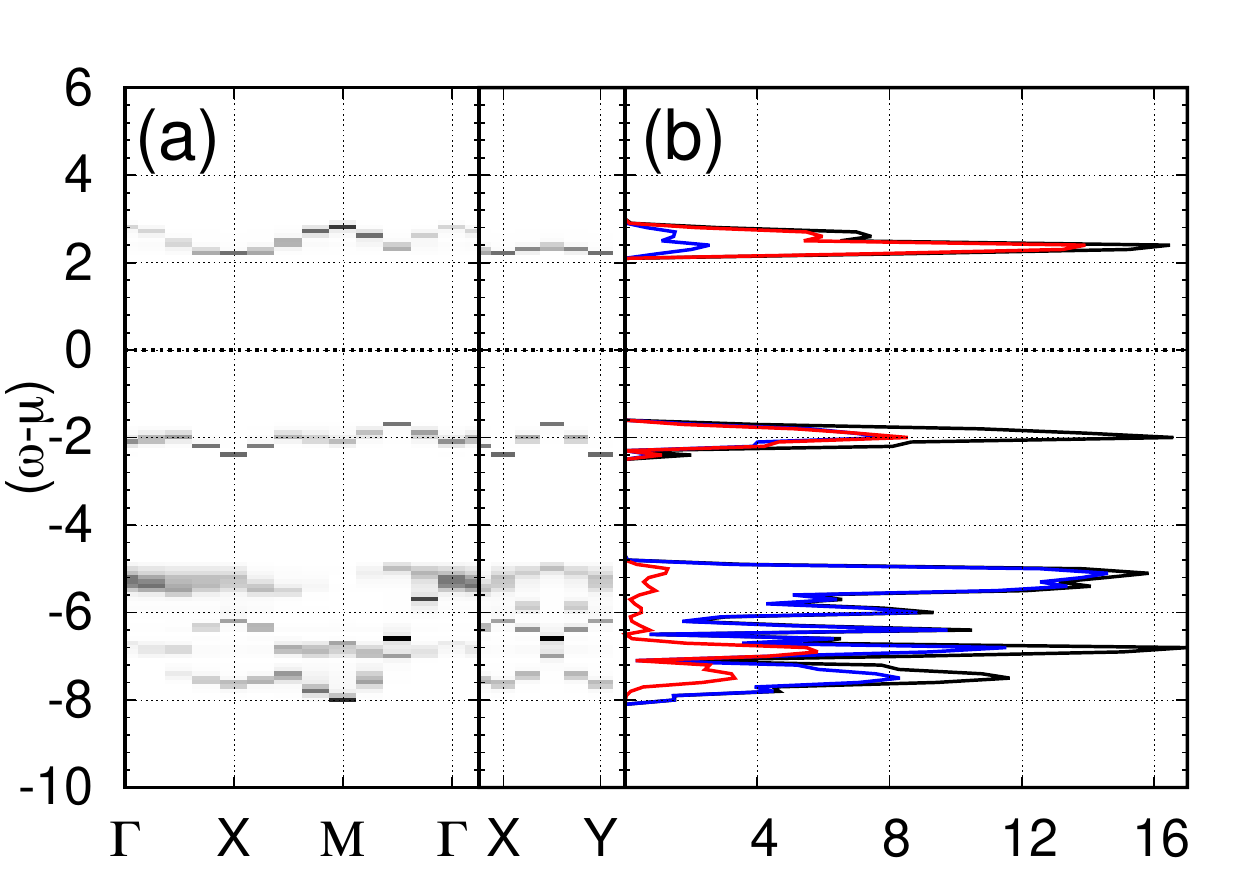}
\vskip -0.3cm
\caption{(color online) (a) Spectral function $A({\bf k},\omega)$ 
along selected directions in the Brillouin zone for 
the spin-fermion model with $J_{\rm Sd}$=3, $J_{\rm Sp}$=1, and $J_{\rm AF}$=0.1 (all in eV) 
using an $8\times 8$ lattice at low temperature $T \sim 30$~K 
in the undoped case. (b) Orbital-resolved
DOS states with parameters as in panel (a).
The orbital spectral weight is 
indicated in red (blue) for the $d$ ($p$) electrons. Black is the total.}
\vskip -0.4cm
\label{dispun}
\end{center}
\end{figure}

\begin{figure}[thbp]
\begin{center}
\includegraphics[width=8cm,clip,angle=0]{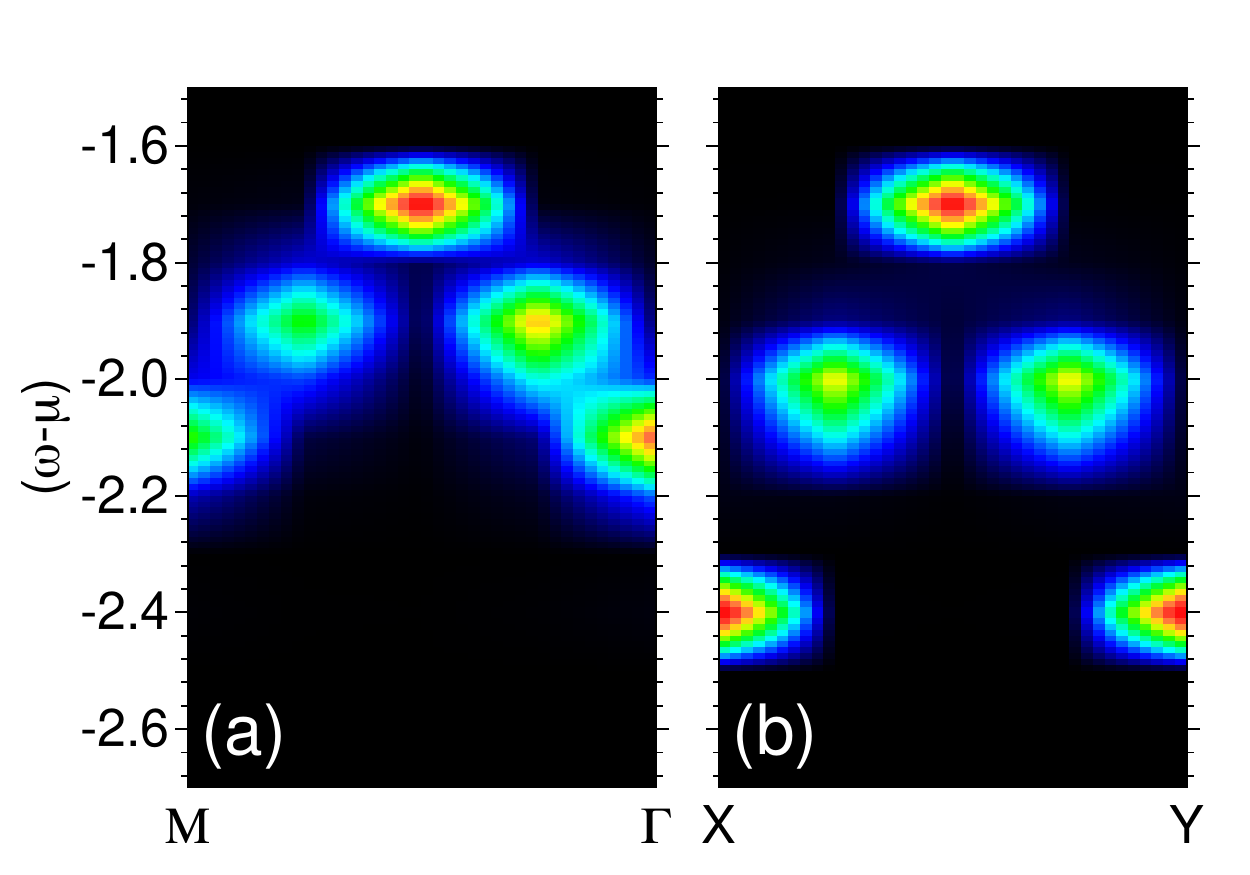}
\vskip -0.3cm
\caption{(color online) Spectral function $A({\bf k},\omega)$ for the ZRS-like 
band using the spin-fermion model with $J_{\rm Sd}$=3, $J_{\rm Sp}$=1, 
and $J_{\rm AF}$=0.1 
(all in eV) on an $8\times 8$ lattice at low temperature $T \sim 30$~K 
in the undoped case. (a) are results along the M-$\Gamma$ direction in the 
Brillouin zone, while (b) is same as (a) but along the X-Y direction.}
\vskip -0.4cm
\label{dispunzr}
\end{center}
\end{figure}

The photoemission one-particle 
spectral functions $A({\bf k},\omega)$ were also calculated and their projection along 
selected directions of the Brillouin zone are shown in 
Fig.~\ref{dispun}~(a) at our lowest temperature $\beta=400$~eV$^{-1}$
(i.e. $T \sim 30$~K). Below the chemical
potential, the closest state in the ZRS-like band is at momentum 
$(\pi/2,\pi/2)$ (half-point in the $M-\Gamma$ and $X-Y$ directions) indicating 
that this will be the momentum of a doped hole, as expected in the cuprates~\cite{elbio,maria}. 
On the other hand, the lowest states in the upper 
band are at $X=(\pi,0)$ and $Y=(0,\pi)$ suggesting that doped 
electrons will have these momenta, as also observed before~\cite{maria}. 
Remarkably, we have found that the maximum around $(\pi/2,\pi/2)$ is 
considerably {\it symmetric} along $\Gamma-M$ and $X-Y$, i.e. with a similar
down curvature, a characteristic 
of the dispersion observed in early photoemission experiments for the undoped Sr$_2$CuO$_2$Cl$_2$ 
cuprate~\cite{wells} that only can be reproduced in one-band Hubbard 
and $t-J$ models by adding diagonal and second nearest-neighbor hoppings~\cite{naza,gooding}. 
In fact, comparing with the experimental data~\cite{wells} the dispersion 
in our results along the directions $\Gamma-M$ and $X-Y$ is 0.5 and 0.8~eV, respectively,
as shown in panels (a) and (b) of Fig.~\ref{dispunzr}, 
close to the 0.3-0.4~eV observed experimentally~\cite{wells}. Note that in the single-band models 
with only nearest-neighbor hoppings the dispersion along $X-Y$ is very 
flat~\cite{wells,naza,gooding} while a stronger dispersion is established 
along that direction in the spin-fermion model because of the $p$-orbitals.

\begin{figure}[thbp]
\begin{center}
\includegraphics[width=8cm,clip,angle=0]{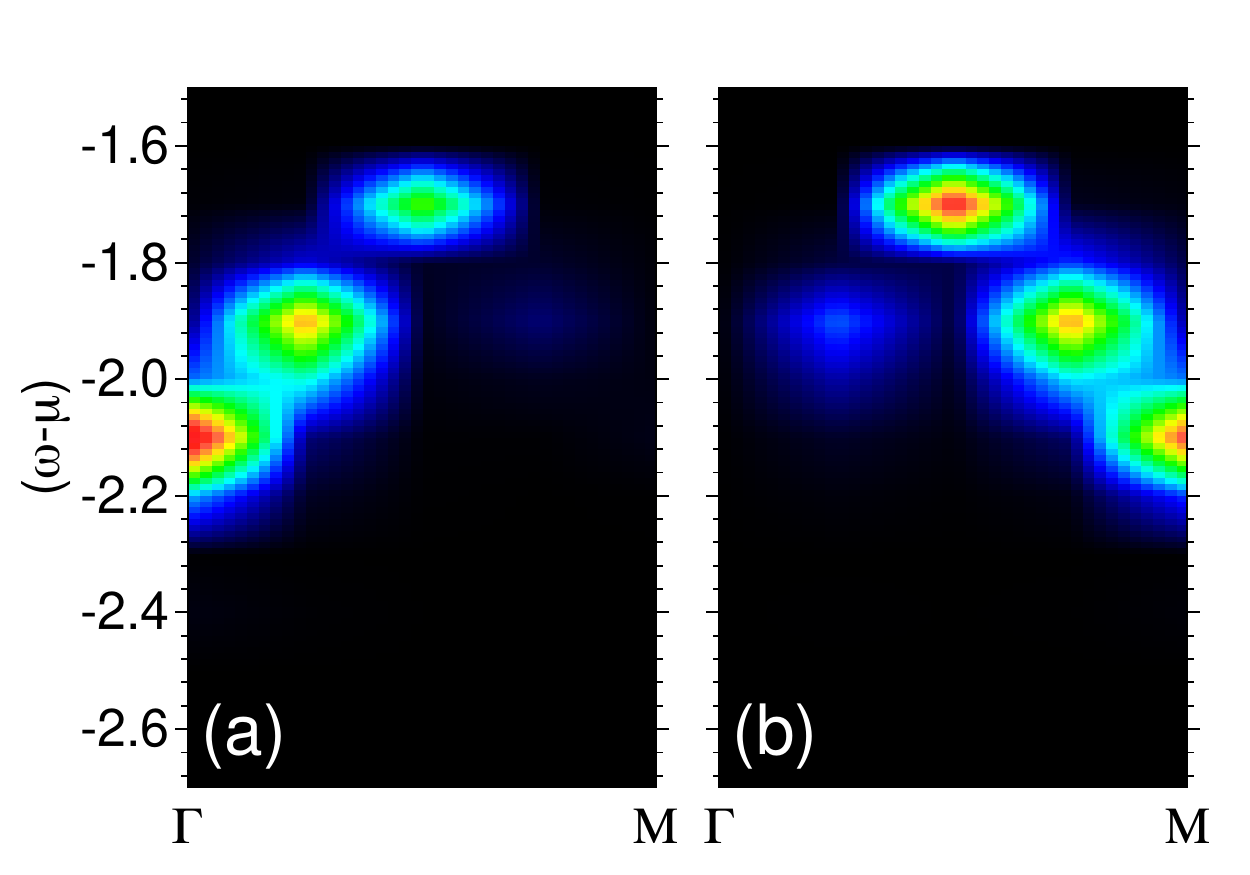}
\vskip -0.3cm
\caption{(color online)
Orbital-resolved spectral function $A({\bf k},\omega)$ for the 
ZRS-like band shown along the $\Gamma$-M direction in the 
Brillouin zone. We use the spin-fermion model with $J_{\rm Sd}$=3, $J_{\rm Sp}$=1, 
and $J_{\rm AF}$=0.1 (all in eV) on an $8\times 8$ lattice at low temperature $T \sim 30$~K
and 
in the undoped case. Panel (a) are results for the $d$-orbital spectral weight, and
(b) for the $p$-orbitals spectral weight.}
\vskip -0.4cm
\label{dispunzrpd}
\end{center}
\end{figure}

The orbital-resolved DOS is displayed in Fig.~\ref{dispun}~(b).  Because the conduction 
band is mostly $d$ in character, doped electrons will be located into $d$-orbitals, 
while the ZRS-like band is a 50-50 mix of $p-d$ character as discussed before. 
This indicates that, due to the additional hybridization 
caused by the interactions, doped holes distribute evenly
among oxygen and copper atoms,
an unusual concept in cuprates where it is widely assumed that holes have entirely oxygen character.
In addition, we have observed that the orbital decomposition supports the identification 
of the charge-transfer band with a ZRS-like band since its $p$-character vanishes 
approaching $\Gamma$, Fig.~\ref{dispunzrpd}~(b), while its $d$-character is small 
close to $M$, see panel (a), in agreement with the phase factor of the ZRS wave 
function~\cite{shastry, eskes, maria}.  
This is similar to results obtained for the three-band Hubbard model~\cite{maria} 
except that the dispersion of the ZRS observed in this previous VCA study 
is of order $t_{pd}\approx 1.3$~eV, slightly larger than the dispersion observed 
experimentally and in the spin-fermion model. 
Finally, in Fig.~\ref{dispun}~(b) there is a lower band, mostly of $p$-character with a small $d$ contribution, similar to the lower spectral weight, 
observed in the three-orbital Hubbard model for $U_d = 8t$ in Fig.~\ref{3bh}~(a).

\begin{figure}[thbp]
\begin{center}
\includegraphics[width=8cm,clip,angle=0]{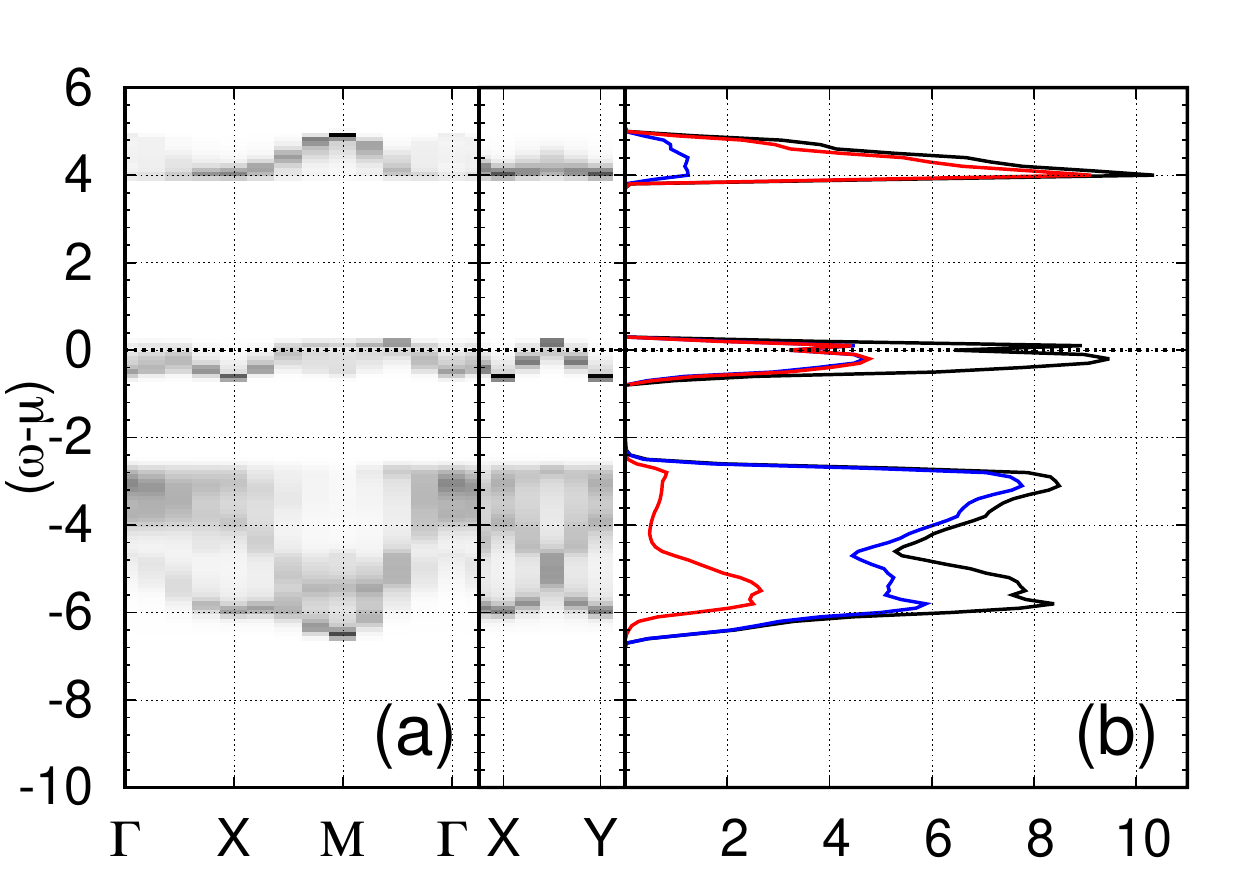}
\vskip -0.3cm
\caption{(color online) (a) Spectral function $A({\bf k},\omega)$ shown along selected directions in the Brillouin zone for 
the spin-fermion model with $J_{\rm Sd}$=3, $J_{\rm Sp}$=1, and $J_{\rm AF}$=0.1 
(all in eV) on an $8\times 8$ lattice,
at low temperature $T \sim 30$~K, and with 25\% hole doping. 
(b) Orbital-resolved DOS
 corresponding to panel (a). The orbital 
spectral weight is indicated in red (blue) for the $d$ ($p$) electrons. Black is the total.}
\vskip -0.4cm
\label{disph16}
\end{center}
\end{figure}

Consider now 25\% hole doping. We focus on this doping 
to compare with results for the three-orbital 
Hubbard model obtained using density functional theory combined with
the dynamical mean-field theory (LDA+DMFT)~\cite{gabi,demedici}. The DOS at different temperatures 
is in Fig.~\ref{dosh0h16}~(b). An important difference with the undoped case [panel (a)] 
is that as the temperature decreases the charge-transfer band develops a pseudogap at 
the chemical potential
(inset of the figure). In Ref.~\cite{gabi} the spectral weight to the right of the chemical potential was identified 
with the quasiparticle, while the spectral weight 
to the left with the incoherent part of the Zhang-Rice singlet. 
Our main features of the DOS  
are in qualitative agreement with those  observed in the LDA-DMFT study of the 
three-orbital Hubbard model: the evolution with doping of the ZRS band shows 
the split of the band into a quasiparticle and an incoherent band. This behavior is 
observed in Fig.~\ref{disph16} along the main directions in the Brillouin zone in panel (a), 
while in (b) the DOS pseudogap develops.

\begin{figure}[thbp]
\begin{center}
\includegraphics[width=8cm,clip,angle=0]{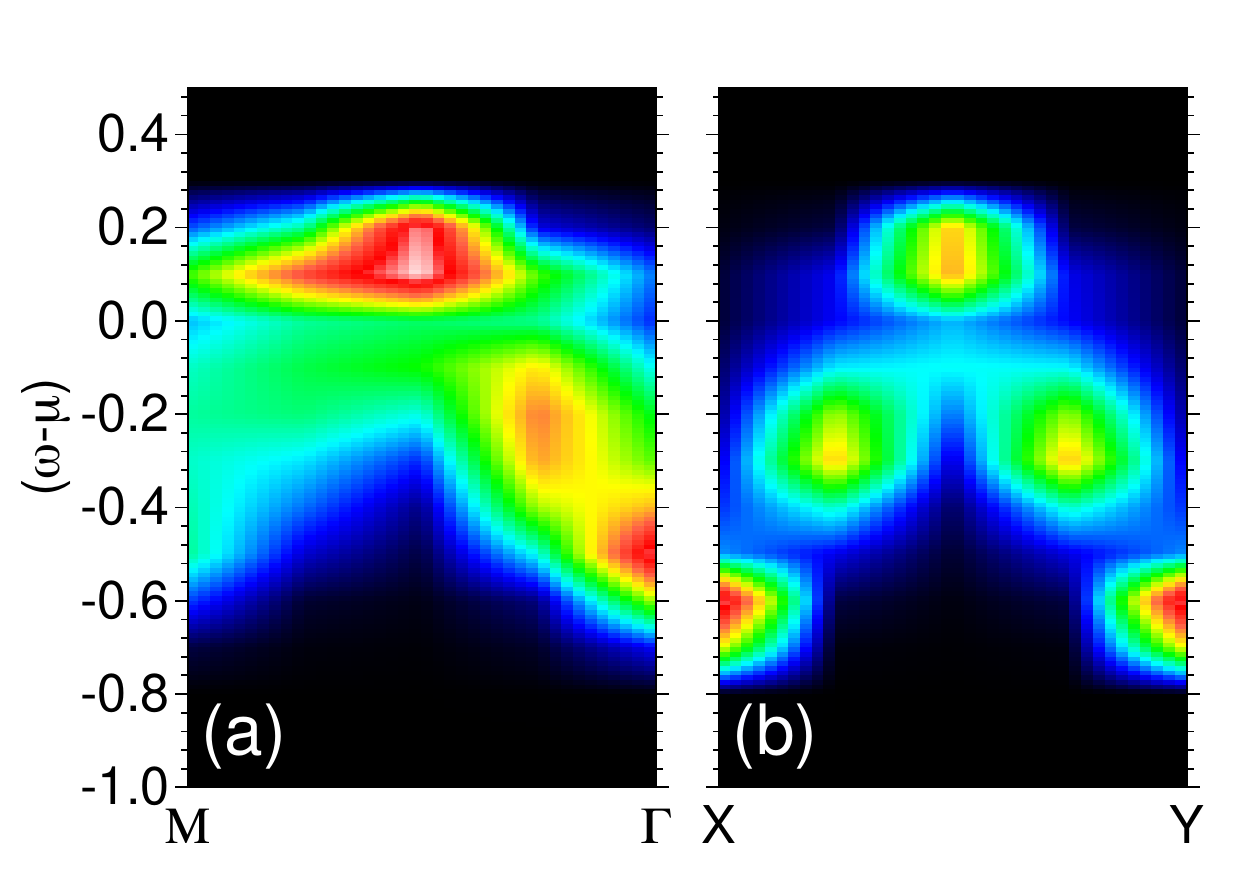}
\vskip -0.3cm
\caption{(color online) (a) Spectral function $A({\bf k},\omega)$ corresponding to the ZRS-like band 
shown along the M-$\Gamma$ direction in the 
Brillouin zone for the spin-fermion model with $J_{\rm Sd}$=3, 
$J_{\rm Sp}$=1, and $J_{\rm AF}$=0.1 (all in eV) using an $8\times 8$ lattice 
at low temperature $T \sim 30$~K and for 16 doped holes. 
(b) same as (a) but for the $X-Y$ direction.}
\vskip -0.4cm
\label{disph16zr}
\end{center}
\end{figure}

Figure~\ref{disph16zr}~(a) shows
that the dispersion is no longer symmetric 
about $(\pi/2,\pi/2)$ along the nodal direction $\Gamma-M$ as in the undoped case.
The quasiparticle peak is below the chemical potential at $\Gamma$ and above at $M$, 
while incoherent weight remains below $\mu$. This feature is crudely reminiscent of the 
``waterfall'' observed experimentally in the cuprates~\cite{lanzara,marshall}. 
In addition, see  panel (b) of the figure, the quasiparticle crosses twice the 
chemical potential along $X-Y$ defining a Fermi surface.

\subsection {Magnetic Properties}

\begin{figure}[thbp]
\begin{center}
\includegraphics[width=8.5cm,clip,angle=0]{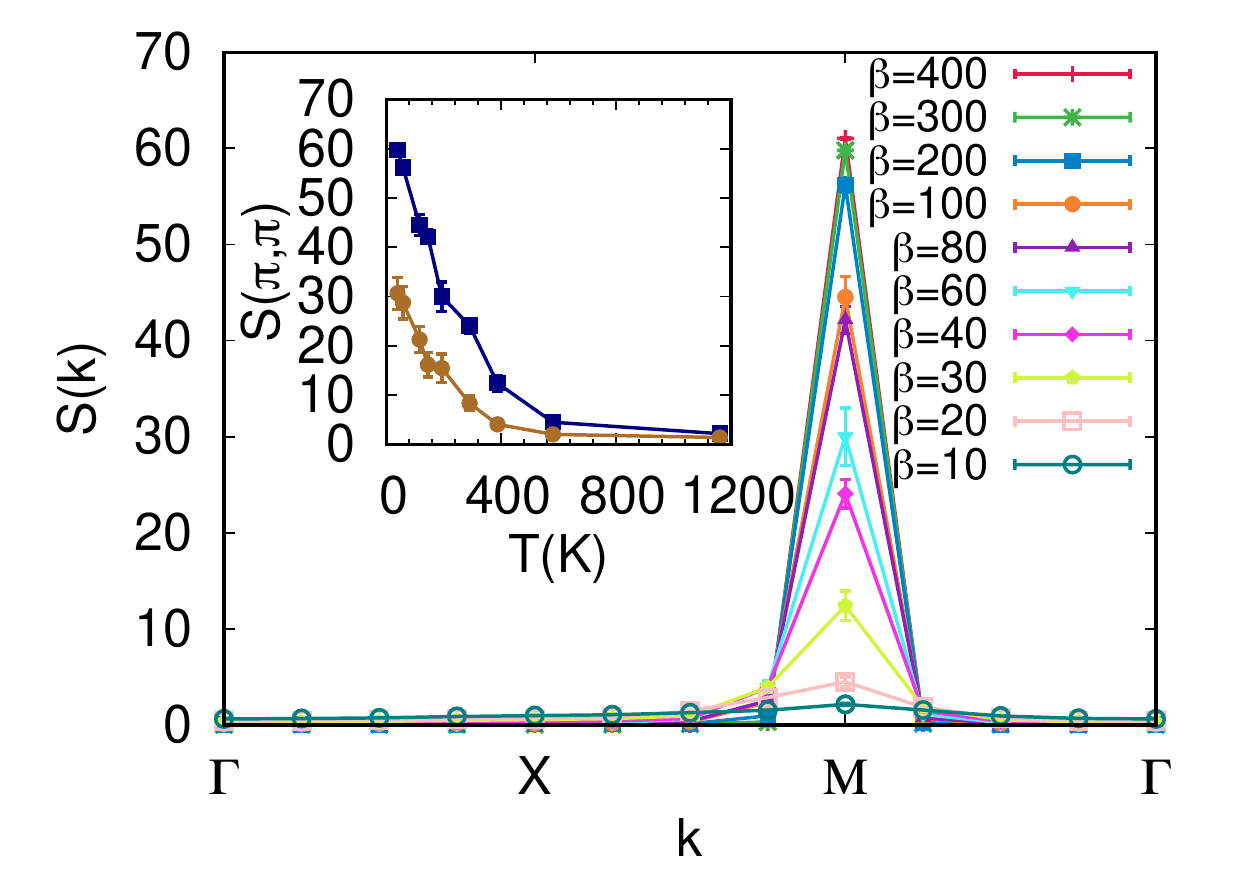}
\vskip -0.3cm
\caption{(color online) Static magnetic structure factor $S({\bf k})$ for 
the localized spins along representative directions in the Brillouin zone 
for the undoped spin-fermion model with $J_{\rm Sd}$=3, $J_{\rm Sp}$=1, 
and $J_{\rm AF}$=0.1 (all in eV) using an $8\times 8$ lattice at various values of 
the inverse temperature ($\beta = 400$ corresponds to $T \sim 30$~K
while $\beta = 10$ to $T \sim 1200$~K). The inset shows 
the evolution of the structure factor at ${\bf k}=(\pi,\pi)$ vs temperature 
for the classical spins (squares) and
for the spins of the electrons in the $d$-orbitals (circles). 
The quantum values have been multiplied by 5 for the sake of comparison with the
results for the classical spins.}
\vskip -0.4cm
\label{skun}
\end{center}
\end{figure}

Consider now the magnetic properties of the model. In the undoped case, 
the system develops long-range antiferromagnetic order in our finite system. 
The real-space spin-spin 
correlation functions between the localized spins are measured versus 
distance and their Fourier transform provides the static magnetic structure factor $S({\bf k})$. 
In Fig.~\ref{skun}, $S({\bf k})$ is shown for various inverse 
temperatures $\beta$ and presented along representative directions in the Brillouin zone. 
The sharp peak is correctly located at $(\pi,\pi)$ and its value increases 
as the temperature decreases, as expected. The inset shows $S({\bf k})$ at ${\bf k}=(\pi,\pi)$ 
varying temperature. A robust antiferromagnetic order starts to develop 
between 200~K and 500~K, in rough quantitative agreement with the real N\'eel temperature 
$T_N\approx$~300~K in the cuprates~\cite{elbio}. In the spin-fermion model, there is a 
natural tendency towards antiferromagnetism due to the nesting of the non-interacting 
Fermi surface, but the addition of a small antiferromagnetic 
Heisenberg coupling $J_{\rm AF}$ between the localized spins further stabilizes 
the expected antiferromagnetic order in the undoped case. The electrons in the Cu 
$d$-orbitals are strongly coupled to the localized spins and their 
spin correlations follow the behavior of the classical spin correlations, as shown 
in the inset of the figure. As a consequence, 
in what follows it is sufficient to focus on the behavior of the classical spins.  

\begin{figure}[thbp]
\begin{center}
\includegraphics[width=8.5cm,clip,angle=0]{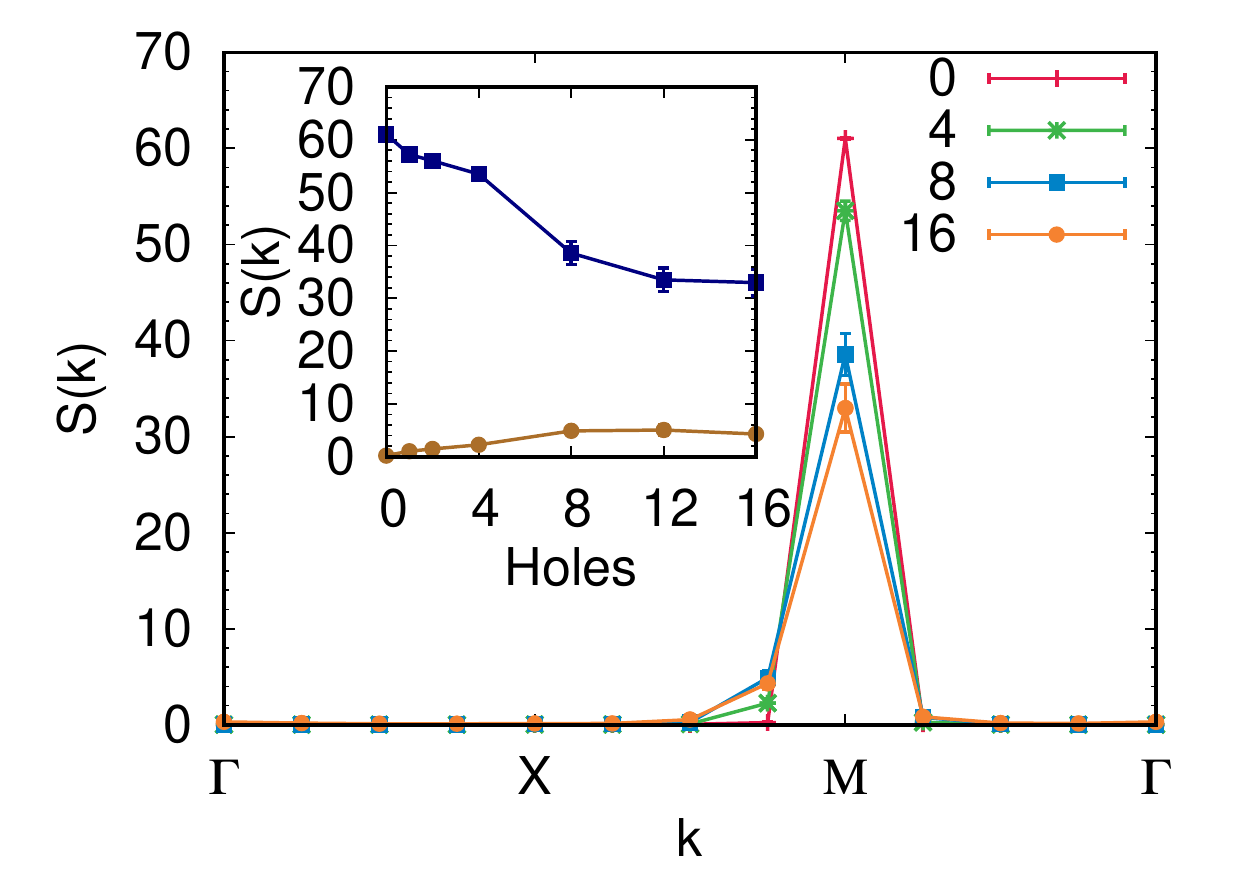}
\vskip -0.3cm
\caption{(color online) Static magnetic structure factor $S({\bf k})$ for 
the localized spins along representative directions in the Brillouin zone 
for the spin-fermion model with $J_{\rm Sd}$=3, $J_{\rm Sp}$=1, 
and $J_{\rm AF}$=0.1 (all in eV) using an $8\times 8$ lattice at temperature 
$T \sim 120$~K and 
for the indicated number of holes. The inset shows the evolution of
$S(\pi,\pi)$ (squares) and $S(\pi,3\pi/4)$ (circles) at $T \sim 120$~K varying 
the number of doped holes as indicated.}
\vskip -0.4cm
\label{sk}
\end{center}
\end{figure}

Upon doping, the antiferromagnetic interaction between the electrons in 
the $p_{\sigma}$ orbitals located at the oxygens and the localized spins 
at the coppers introduces magnetic frustration. This slightly affects the antiferromagnetic order 
as observed in the curves for different dopings in Fig.~\ref{sk}. 
The intensity of the peak at ${\bf k}=(\pi,\pi)$ decreases, while the weight of $S({\bf k})$
at ${\bf k}=(\pi,3\pi/4)$ increases, as shown 
in Fig.~\ref{sk} where $S({\bf k})$ is presented at temperature $T \sim 120$~K 
along representative directions in the Brillouin 
zone for different electronic densities. The increasing transference of 
weight to $(\pi,3\pi/4)$ with hole doping captures qualitatively the expected 
trend towards the well-known magnetic 
incommensurability of the cuprates~\cite{cheong,pdai,mook} at momentum 
$(\pi,\pi-\delta)$ and $(\pi-\delta,\pi)$. Experimental evidence has indicated 
that this incommensurability is related to stripe structures either static or 
dynamical~\cite{tranquada} and more recently the possibility of states with 
intertwined spin, charge, and superconducting orders was also proposed~\cite{berg}.
The study of the possible existence of stripes, ZRS structures, 
high-spin polarons, and intertwined states in the ground state upon doping 
are future projects that can be addressed via the three-orbital spin-fermion model 
introduced here.

\section{Conclusions}\label{conclu}

In this publication, a phenomenological three-orbital model that reproduces the charge-transfer properties of 
superconducting cuprates was introduced. The notorious 
difficulty to incorporate the electronic Coulomb repulsion
of the multiorbital Hubbard model was alleviated by introducing 
antiferromagnetic interactions between the spins of the electrons
in the three itinerant orbitals and phenomenological spins located at the coppers. 
The interaction of the $d$-electrons
with the localized spins effectively induces a gap in the half-filled $d$-band and prevents double occupancy, similarly
as the Hund interaction does in double-exchange models for manganites.
Considering the localized spins as classical, 
as in similar models for  manganites~\cite{manganites}, one-orbital cuprates~\cite{opstripes}, 
and iron-based superconductors~\cite{shuhua13}, the Hamiltonian
becomes quadratic in the fermionic fields and it can be studied by classical 
Monte Carlo
combined with the diagonalization of the effective single-particle quantum Hamiltonian. 
This process allows the study of a three-orbital
model in larger clusters  than those
 accessible to multi-orbital Hubbard models and, moreover, the full range of 
temperatures can be explored. 

Several features of the band structure experimentally observed in the cuprates 
are well-reproduced by this simplified new model, such as the development of
a charge-transfer gap in the undoped case framed by a conduction band of mostly 
$d$-character with minima at momentum $(\pi,0)$ and $(0,\pi)$ and a
ZRS-like band with a 50/50 contribution from $p$ and $d$ orbitals with a maximum at $(\pi/2,\pi/2)$. 
In addition, the band dispersion about the maximum is symmetric
along $\Gamma-M$ and $X-Y$ in the Brillouin zone, an experimental property of the 
cuprates that is not captured by single orbital models unless
$t'$ and $t''$ hoppings are added. Upon doping, a pseudogap in the ZRS band develops 
at the chemical potential and spectral features resembling the ``waterfall'' are observed.

The correct magnetic properties are also captured by the spin-fermion model 
that displays clear tendencies towards long-range antiferromagnetic order
in the undoped case. It also starts to show incipient indications of incommensurability 
along $(\pi-\delta,\pi)$ and $(\pi,\pi-\delta)$ in the doped case. These features upon doping,
which may originate in stripes or intertwinned order and that may require cylindrical boundary
conditions for their stabilization,
can only be seen clearly using larger clusters and they will be the subject of future work.

\section{Acknowledgments}

Discussions with C.B. Bishop are acknowledged.
E.D. and A.M. were supported by the US Department of Energy, 
Office of Basic Energy Sciences, Materials Sciences and Engineering
Division. M.H. was supported by the National Science Foundation, under
Grant No. DMR-1404375. M.D. was supported by the Deutsche Forschungsgemeinschaft,
via the Emmy-Noether program (DA 1235/1-1) and FOR1807 (DA 1235/5-1).



\begin{references}
\bibitem{nohyb} Notice that $U_d\gg 16t$ is needed in order to destroy the $p-d$ hybridization.
\bibitem{negdel} In a future publication we will explore the case $\epsilon_p=3.6$~eV 
with the $p$ band above the $d$ one, corresponding to a ``negative charge-transfer gap'' 
($\Delta<0$), following the choice in Ref.~\cite{fratino,demedici} to emphasize 
the charge-transfer properties of the system ensuring that the doped holes go 
into almost pure $p$ orbitals.
\bibitem{foot1} The temperature $T$ in Kelvin degrees is obtained using $T=11,605/\beta$.
\end{references}
\end{document}